\documentclass[a4paper]{article}

\usepackage[backend=biber,style=numeric,sorting=none]{biblatex}
\usepackage[top=3cm,bottom=3.5cm,left=2.5cm,right=2.5cm]{geometry}
\usepackage[utf8]{inputenc}
\usepackage{graphicx}
\usepackage{amsfonts,amssymb,amsmath}
\usepackage{hyperref}
\hypersetup{%
	pdfauthor={Giulia Maniccia and Giovanni Montani},%
	pdftitle={Quantum Gravity Corrections to the inflationary spectrum in a Bohmian approach}%
}

\usepackage{mathrsfs}

\newcommand{\ampl}{\sqrt{\rho}}
\newcommand{\phase}{e^{\frac{i}{\hbar}S}}
\newcommand{\dev}[1]{\frac{\partial #1}{\partial v}}
\newcommand{\devsq}[1]{\frac{\partial^2 #1}{\partial v^2}}

\newcommand{\deusq}[1]{\partial_{\varphi_{\mathbf{k}} }^2 #1 }

\newcommand{\Hubble}{\mathscr{H}}
\newcommand{\uscal}{\varphi_{\mathbf{k}}}

\newcommand{\myvec}[1]{\mathbf{#1}}

\title{Quantum Gravity Corrections to the inflationary spectrum in a Bohmian approach}
\author{%
{Giulia Maniccia}\textsuperscript{1,2}\thanks{giulia.maniccia@uniroma1.it} \,and Giovanni Montani\textsuperscript{3,1}\thanks{giovanni.montani@enea.it}}

\date{
\footnotesize\textsuperscript{\textbf{1}}Physics Department, “La Sapienza” University of Rome, P.le A. Moro 5, 00185 Roma, Italy\\ \textsuperscript{\textbf{2}}Department Physik, Institut für Quantengravitation, Theoretische Physik III, Friedrich-Alexander
Universit\"{a}t Erlangen-N\"urnberg, Staudtstr. 7/B2, 91058 Erlangen, Germany\\
\textsuperscript{\textbf{3}}FNS Department, ENEA, C.R. Frascati, Via E. Fermi 45, 00044 Frascati, Italy }

\begin{document}

\maketitle

\abstract{
A precise interpretation of the universe wave function is forbidden in the spirit of the Copenhagen School since a precise notion of measure operation cannot be satisfactorily defined.
Here, we propose a Bohmian interpretation of the isotropic universe quantum dynamics, in which the Hamilton--Jacobi equation is restated by including quantum corrections, which lead to a classical trajectory containing effects of order $\hbar^2$. This solution is then used to determine the spectrum of gauge-invariant quantum fluctuations living on the obtained background model. The analysis is performed adopting the wave function approach to describe the fluctuation dynamics, which gives a time-dependent harmonic oscillator for each Fourier mode and whose frequency is affected by the $\hbar^2$ corrections. The properties of the emerging spectrum are discussed, outlining the modification induced with respect to the scale-invariant result, and the hierarchy of the spectral index running is discussed.
}


\maketitle

\section{Introduction}

The Copenhagen School interpretation of Quantum Mechanics (QM) \cite{bib:QMinterpretations} fully succeeded in interpreting the fundamental experiments on atomic and molecular spectra and it remained, over~the years, the~privileged point of view, providing a viable phenomenology for quantum field theory too~\cite{bib:bjorken-drell-rqm}.
However, no rigorous and unique proof exists that the Copenhagen School formulation of the ``measurement operation'' is the only viable description for the quantum phenomena. 
In the 1950s, de Broglie and Bohm~\cite{bib:bohm-1952,bib:debroglie-1958} proposed a completely different approach to the description of quantum phenomenology, based on the idea that the indeterminism results into a ``quantum potential'' acting on a classical trajectory, in~close analogy to a stochastic mechanical approach~\cite{bib:smolin-1986}.
This alternative formulation, which deprives the physical (quasi-classical) observer from its central role in the phenomenology of a quantum dynamics, turns out to be of particular interest in those situations in which the definition of such an observer is ambiguous, if~not even impossible, like in quantum gravity physics~\cite{bib:thiemann-book}. 

Among these special physical contexts, Quantum Cosmology stands as the most appropriate arena in which to implement the de Broglie--Bohm (dBB) interpretation.
In fact, in Quantum Cosmology reviews~\cite{bib:halliwell-review-1989,bib:kiefer-1994-review,bib:montani-benini-lunga-2008,bib:montani-barca-giovannetti-2021,bib:peter-kiefer-2022,bib:maniccia-deangelis-2022}, the~difficulty of defining a quasi-classical observer appears evident when the whole universe is subjected to a quantum dynamical regime, which~we expect happened in the Planckian era~\cite{bib:isham-book-1993,bib:montani-primordialcosmology,bib:montani-cqg}.
A different situation is realized with a separation of the system between a background quasi-classical cosmology and a fully quantum subsystem~\cite{bib:rubakov-lapchinsky-1979,bib:vilenkin-1989,bib:barvinsky-1990,bib:bertoni-venturi-1996,bib:deangelis-montani-2020,bib:maniccia-montani-antonini-2023}. 
However, also in these hybrid situations, the~concept of the observer as well as of the measurement operation suffer from non-trivial formal ambiguities~\cite{bib:gundhi-bassi-2021}.
Thus, in~Quantum Cosmology, the~dBB formulation appears as a natural interpretative tool to extract phenomenological information from the canonical quantization as~described via a ``quantum trajectory'' instead of the standard universe wave function~\cite{bib:blyth-isham-1975,bib:hartle-hawking-1983}.

Here, we address exactly this point of view, having the scope of calculating the quantum gravity corrections to the inflationary primordial spectrum. Thus, we start from a standard Wheeler--DeWitt (WDW) equation for an isotropic universe in the presence of a cosmological constant, to~be thought as an inflationary potential in the slow-rolling region~\cite{bib:kolb-turner,bib:montani-primordialcosmology,bib:peter-uzan-PC,bib:kirillov-montani-2002,bib:cianfrani-montani-muccino-2010}. 
Then, we perform the separation of the universe wave function in its modulus and phase, respectively, searching for the dynamical equations of the imaginary and real parts of the corresponding Hamiltonian constraint. 
As a result, the~real part turns out to be a modified
Hamilton--Jacobi for the de Sitter universe, containing a quantum potential which is due to the non-constant modulus of the wave function.
The imaginary part links this modulus to the phase of the wave function, whose dependence can be explicitly expressed.
Thus, we arrive at a single equation for the phase, corresponding to a Hamilton--Jacobi formulation, amended for $\hbar^2$-corrections.
Solving this equation and restoring the standard relation between the momentum and the phase derivative, we arrive at a modified classical dynamics of the de Sitter universe in the presence of $\hbar^2$ corrections, coming from quantum cosmological
features.

Then, we use this revised ``classical'' background to study the inflationary spectrum as being due to the quantum fluctuations of a free massless scalar field, i.e.,~the same inflaton responsible for the vacuum energy which
exponentially expands the universe. 
According to a Born--Oppenheimer separation of the dynamics~\cite{bib:bertoni-venturi-1996,bib:venturi-2017,bib:venturi-2021,bib:maniccia-montani-2022} and working at the dominant order in $\hbar^2$, we study the quantum fluctuations of the inflaton field and solve the associated time-dependent harmonic oscillator, describing its Fourier components dynamics. 
As a fundamental result of this analysis, we calculate
a revised inflationary spectrum which contains quantum gravity corrections up to order $\hbar^2$ and which is the dBB counterpart of the same study performed via a standard canonical formulation in~\cite{bib:brizuela-kiefer-2016-desitter,bib:brizuela-kiefer-2016-slow-roll,bib:venturi-2020,bib:maniccia-montani-torcellini-2023}.

The manuscript is structured as follows. In~Section~\ref{sec:intro-dbb}, we review the main points of the dBB interpretation of Quantum Mechanics; such an approach is then applied to the background cosmological picture in Section~\ref{sec:dbb-background}. Section~\ref{sec:dbb-bo} is devoted to the setting of the Born--Oppenheimer approximation for the model consisting of the inflaton field evolution over such a dBB background. In~Section~\ref{sec:applic-spectrum}, the computation of the primordial perturbations spectrum is carried out. Final remarks and outlooks are presented in Section~\ref{sec:conclusions}.

\section{Overview of the de Broglie-Bohm approach}\label{sec:intro-dbb}

The de Broglie--Bohm approach, also referred to as pilot wave theory~\cite{bib:bohm-1952,bib:debroglie-1958,bib:holland-1993-dbb}, provides a deterministic quantum theory alternative to QM.
In this framework, the~core elements of the quantum theory are postulated to exist independently of observation or measurement (for the gravitational case, the~geometry of the 3d hypersurfaces and their canonical momenta identified via the ADM representation~\cite{bib:mtw-gravitation}), {without any need for the wave function collapse.}
The evolution of the elements of the theory diverges from classical dynamics due to the emergence of a quantum potential; however, the~deterministic character is assured by the guiding wave function, which uniquely identifies the evolution of particles and their positions in space, with~suitable initial~conditions.

{For a comprehensive review on this topic, including its historical development, explanation of the measurement process, and applications, we refer to the reviews~\cite{bib:oriols-2019-bohmian,bib:pintoneto2019bohmian}.}
A simple case to illustrate this approach is the non-relativistic particle, whose quantum dynamics is expressed by the Schr\"odinger equation
\begin{equation}\label{eq:7-schrod-1particle}
    i\hbar \partial_t \Psi (x,t) =\left( -\frac{\hbar^2}{2m}\partial_x^2 + V(x) \right) \Psi(x,t)\,.
\end{equation}

{Here,} we have the freedom to rewrite the complex wave function as 
\begin{equation}\label{eq:7-dbb-ansatz}
    \Psi (x,t)= A(x,t) \,e^{\frac{i}{\hbar}S(x,t)}
\end{equation}
where $A$ and $S$ are both real. We stress the presence of a formal analogy between \eqref{eq:7-dbb-ansatz} and the ansatz of~\cite{bib:vilenkin-1989}, i.e.,~Vilenkin's proposal for the identification of a quantum subsystem in the semiclassical regime. However, here, such a form of the wave function carries a different connotation: in~\cite{bib:vilenkin-1989}, such writing is necessary to perform the semiclassical WKB expansion and reconstruct a time parameter from the scalar WDW constraint. Here, instead, we already start from a full Schr\"odinger dynamics in the label time. 
In this sense, it is important to remark that the dBB interpretation relies on the idea of small quantum corrections i.e.,~it identifies the $\hbar$ parameter as small, while keeping the whole set of constraints; on the other hand, the~WKB method effectively reduces the system's equations via an $\hbar$ expansion, essentially providing physically meaningful results in the limit $\hbar \rightarrow 0$.
The similarity between the two ansatzs will be further discussed in relation to the cosmological picture considered in Section~\ref{sec:dbb-bo}.

The expression~\eqref{eq:7-dbb-ansatz} automatically decomposes the dynamics~\eqref{eq:7-schrod-1particle} into two partial differential equations, i.e.,~its real and imaginary parts, respectively:
\begin{gather}
    \frac{\partial S}{\partial t} + \frac{(\partial_x S)^2}{2m} + V - \frac{\hbar^2}{2m}\frac{\partial_x^2 A}{A}=0\,,\label{eq:7-eq-phase}\\
    \partial_t A^2 +\frac{1}{m} \partial_x \left( A^2 \partial_x S\right) =0\,. \label{eq:7-eq-amplitude}    
\end{gather}
{In} the Copenhagen interpretation of QM, Equation~\eqref{eq:7-eq-amplitude} represents the continuity equation for the probability density $A^2$ associated to finding the particle at a given position $x$ in time. In~this sense, all physical content is given by this equation so that the phase and Equation~\eqref{eq:7-eq-phase} turn out to be superfluous for a given~state.

The dBB interpretation provides instead a larger view of both components. In~particular, Equation~\eqref{eq:7-eq-phase} can be recognized as the Hamilton--Jacobi equation plus an extra term, labeled as the \emph{{quantum potential}}:
\begin{equation}
    Q(x,t) = -\frac{\hbar^2}{2m} \frac{\partial_x^2 A}{A}(x,t)\,.
\end{equation}
Here, we observe that $Q$ depends only on the
form of $\Psi$, i.e.,~on its amplitude $A$.
This potential acts as a quantum effect, modifying the  phase dynamics from its classical form governed by the Hamilton--Jacobi (HJ) equation; see Equation~\eqref{eq:7-eq-phase}.
Therefore, the~particle trajectory determined by the solution of the \emph{{guidance equation}}
\begin{equation}\label{eq:7-guidance-eq}
    p = m \dot{x} = \partial_x S(x,t)\,
\end{equation}
takes into account such quantum effects due to the presence of $Q$. 
Clearly, the~classical limit is recovered in the regime $Q\rightarrow 0$ so that Equation~\eqref{eq:7-eq-phase} reduces to the standard HJ equation and the physical predictions coincide with the classical behavior. 
In this case, the set of Equations~\eqref{eq:7-eq-phase} and \eqref{eq:7-eq-amplitude} can be interpreted as describing an ensemble of classical particles under the influence of a classical potential $V$, with~velocity field $\partial_x S/m$.

The dBB interpretation also allows for a simple explanation of the measurement process: observational outcomes are determined by the interactions between the guiding wave function and the measuring apparatus, which preserve the unitary evolution of the quantum state; in other words, no collapse of the wave function is~necessary. 

The description of quantum phenomena through definite particle trajectories as~facilitated by such a framework opens possibilities for novel methodologies in handling multi-particle systems (for a generalization to many-particle settings, see~\cite{bib:oriols-2019-bohmian}), such as in quantum hydrodynamics (e.g., \cite{bib:wyatt-2000}).
{In the context of Quantum Cosmology, this interpretation opens new perspectives such as the spacetime singularity resolution (see~\cite{bib:pintoneto2019bohmian}); here, we will focus on its application to address the description of quantum perturbations of the gravitational background and their influence on the inflaton sector.}
Nonetheless, it shall be remarked that experimental verifications of the dBB predictions may present significant challenges~\cite{bib:holland-1993-dbb}.

\section{The Background dBB~Picture}\label{sec:dbb-background}

It is interesting to consider the dBB picture in the gravitational context. Such an approach has been shown to influence the constraint algebra in quantum gravity and the dynamical sector of Einstein's equations~\cite{bib:shojai-2003}; in black hole systems, the~Hawking radiation has been predicted to stop before total evaporation due to the quantum trajectories modifying the Schwarzschild setting~\cite{bib:das-BHdbb-2014,bib:ali-BHdbb-2016}.

For Quantum Cosmology, the~dBB approach has been implemented as a tool to investigate the primordial evolution of the universe~\cite{bib:glikman-1990,bib:vink-1992,bib:shtanov-1996,bib:pinto-neto-2013,bib:peter-2018} and its boundary conditions~\cite{bib:pintoneto-1999-boundary}; a remarkable result from such dBB trajectories is the prediction of a universe avoiding the primordial singularity, usually through a quantum bounce~\cite{bib:colistete-pintoneto-1998,bib:acacio-pintoneto-1998,bib:peter-2007,bib:peter-vitenti-2016,bib:pinto-neto-2020-dbb,bib:zampeli-2021,bib:malkewicz-peter-2022,bib:molinari-pintoneto-2024}.

In the present paper, we compute the dBB universe volume trajectory in order to analyze the consequences on the primordial inflaton perturbations, which are superimposed on a pure de Sitter background expressed through a simple minisuperspace model~\cite{bib:montani-primordialcosmology,bib:kiefer-QG}.

\subsection{The de Sitter Expanding~Universe}\label{ssec:desitter-dbb}

We consider a spatially flat isotropic universe described by the FLRW line element:
\begin{equation}
        ds^2 = -N^2 dt^2 + a^2(t) (dx^2 +dy^2 +dz^2)\,.
    \end{equation}
{On} such a background, we will introduce small perturbations of the scalar inflaton field $\phi$; see Section~\ref{sec:dbb-bo}.
In accordance with the homogeneity of the model, we can reduce the full minisuperspace formalism to the two minisuperspace variables, associated to the scale factor and scalar inflaton~fluctuations.

At the action level, we insert a positive cosmological constant in order to put ourselves in the (pure) de Sitter regime. Indeed, the~exponential expansion of inflation is propelled by the inflaton field rolling down its potential $V(\phi)$; in the regime of nearly constant potential, a~suitable description is provided by replacing $V(\phi)\simeq const.$ with an effective (positive) cosmological constant contribution $\Lambda$. 
This reflects onto the expansion rate of the universe as $\bar{\Hubble} \propto \sqrt{\Lambda}$, i.e.,~the Hubble parameter $\Hubble = \dot{a}/a$ takes a constant value, giving the classical solution $a =e^{\bar{\Hubble} t}$ (being $t$ the label time). 

The super-Hamiltonian takes the form (n $c=1$ units):
\begin{equation}
H = -\frac{1}{48 M} \frac{p_a^2}{a}  + 4M \Lambda a^3 +H^{\phi}\,,
\end{equation}
where $M = 1/(32\pi G) = m_\mathrm{P}^2/(4 \hbar)$, where $m_{\mathrm{P}}$ is the reduced Planck mass, {and $p_a$ is the momentum canonically conjugate to $a$, corresponding to $p_a= -24M\, a\, \dot{a}$ from the Hamilton equation~\cite{bib:rubakov-book}}.
Here, the scalar inflaton field $\phi$ represents the only matter content; it is precisely the small fluctuations of $\phi$ on such a background that result in the power spectrum being investigated.
For convenience of the analysis, we make a canonical transformation to use a square-root volume variable $v=a^{3/2}$ instead of the scale factor $a$, obtaining the following super-Hamiltonian for the gravitational background (indicated by the label $bk$): 
\begin{equation}\label{eq:7-applic-WDW-v}
H^{bk} = -\frac{3}{64 M} p_v^2  + 4M \Lambda v^2\,.
\end{equation}

{{For} an application of the super-Hamiltonian formalism in a different research area, see~\cite{bib:valiente-2019}.}

\subsection{Modified Trajectory for the Gravitational~Background}\label{ssec:background-mod-trajectory}

Following  Section~\ref{sec:intro-dbb}, we now wish to derive the dBB trajectory for the scale factor (actually for the variable $v=a^{2/3}$) of the universe in this minisuperspace~model.

We write the Bohmian wave function \eqref{eq:7-dbb-ansatz} as 
\begin{equation}\label{eq:separation-psi-rho-s}
\psi(v) = \sqrt{\rho(v)} \;e^{\frac{i}{\hbar}S(v)}\,,
\end{equation}
which refers to the background only; both $\sqrt{\rho(v)}$ and $S(v)$ are real, and~the ``quantum'' character is encoded in the phase~contribution.

Although the trajectory is described by the guidance Equation \eqref{eq:7-guidance-eq} and depends on the phase $S$ only, the~dynamics of $\ampl$ and $S$ are actually coupled as~can be inferred by the real and imaginary parts of the quantized gravitational constraint. Indeed, considering the background WDW equation
\begin{equation}\label{eq:wdw}
    \left(\frac{3\hbar^2 }{64 M} 
   \devsq{} + 4M \Lambda v^2 \right) \psi (v) =0
\end{equation}
we find the following real and imaginary contributions, labeled with $\mathcal{C}_{\Re}$ and $\mathcal{C}_{\Im}$, respectively:
\begin{gather}
    \mathcal{C}_{\Re} := \frac{3}{64M} \hbar^2 \devsq{\ampl} -\frac{3}{64M} \ampl \left(\dev{S}\right)^2 +4M \Lambda v^2 \ampl =0\,, \label{eq:7-applic-real-constr}\\
    \mathcal{C}_{\Im} := \frac{3}{64M} i\hbar \left( 2\dev{\ampl} \dev{S} + \ampl\, \devsq{S} \right) = 0\,.\label{eq:7-applic-im-constr}
\end{gather}

Actually, we remark that that the Wheeler--DeWitt constraint associated to a canonically quantized system can sometimes be solved exactly:
for instance, Equation~\eqref{eq:wdw} admits a solution in terms of the Bessel functions (see~\cite{bib:cianfrani-montani-muccino-2010}).
However, the reconstruction {a posteriori} of the phase and modulus of the wave function from a solution of the WDW equation does not in general commute with solving the coupled equations for these two same quantities, both from a conceptual and technical point of view.
While the starting WDW Equation~\eqref{eq:wdw} contains linear differential operators only, Equation~\eqref{eq:7-applic-real-constr} is clearly non-linear in the Hamilton--Jacobi component; this in principle prevents a direct mapping of the space of solutions for the two setups.
The structure of the WDW constraint can provide different solutions in distinct regions of the quantum numbers involved in the problem (see~\cite{bib:cianfrani-montani-muccino-2010}), a~feature with an unclear correspondence in the dBB scheme.
In the dBB approach, the~physical content of the theory is dictated by the setup of a coupled system of equations for the amplitude and phase. 
Indeed, the emergence of a generalized HJ function is the basis of the idea that quantum physics is summarized by the additional quantum potential, while the original (classical) concept of trajectories is still preserved.
Therefore, we here address {ab~initio} the analysis of Equation~\eqref{eq:7-applic-im-constr} for the wave functional phase coupled with Equation~\eqref{eq:7-applic-real-constr}, disregarding any memory of the original Wheeler--DeWitt formulation.

Consistently with the Bohmian interpretation,  $\mathcal{C}_{\Re}$ \eqref{eq:7-applic-real-constr} corresponds in the $\hbar \rightarrow 0$ limit to the standard HJ equation 
\begin{equation}\label{eq:7-appl-HJ}
    \ampl \left[ -\frac{3}{64M} \left(\dev{S}\right)^2 + 4M \Lambda v^2 \right]=0\,.
\end{equation}
{Indeed,} the first term of Equation~\eqref{eq:7-applic-real-constr}, which is of order $\hbar^2$, is the quantum potential giving the deviation from the purely classical solution. 
{We recall that, for~the HJ theory on curved spacetime, many investigations have been carried out in the context of cosmology; see, for example,~\cite{bib:rahman-2012,bib:ilias-2013,bib:sakalli-2013} for black hole radiation. Here, however, we will focus on the modified HJ equation emerging from \eqref{eq:7-applic-real-constr} in the dBB picture.}

Now focusing on the imaginary part of the constraint, we recognize that Equation~\eqref{eq:7-applic-im-constr} can be rewritten as $ \partial_v \left[(\ampl)^2 \partial_v S \right] /\ampl$ so that we can immediately solve the amplitude in terms of the phase:
\begin{equation}\label{eq:7-rho-function-s}
    \ampl = \frac{c}{|\partial_v S|^{1/2}}
\end{equation}
where $c$ is a numerical constant. 
Substituting into \eqref{eq:7-applic-real-constr} and dividing by $\ampl$, one is led to the following equation for $S(v)$:
\begin{equation}\label{eq:7-eq-exact-S}
    -\frac{3}{64 M} (\partial_v S)^2 + 4 M \Lambda v^2 =-\frac{9\hbar^2}{256 M} \frac{(\partial_v^2 S)^2}{(\partial_v S)^2} +\frac{3\hbar^2}{128 M} \frac{\partial_v^3 S}{\partial_v S}\,,
\end{equation}
which is a third-order non-linear inhomogeneous differential equation, of~which a closed form is not found.
Since we wish to consider the small deviations from the classical behavior, attributed to the quantum potential and responsible for the right-hand side of \eqref{eq:7-eq-exact-S}, we look for an approximate solution at the leading order:
\begin{equation}\label{eq:7-applic-exp-S}
    S(v) = S_0 (v) +\hbar^2 S_1(v)\,.
\end{equation}
{This} expression can be interpreted as the WKB expansion of the gravitational phase alone in \eqref{eq:7-dbb-ansatz}; see~\cite{bib:vilenkin-1989,bib:montani-digioia-maniccia-2021}. Clearly, $S_0$ must correspond to the (classical) Hamilton--Jacobi solution of \eqref{eq:7-appl-HJ}, i.e.,~for an expanding universe
\begin{equation}\label{eq:7-appl-sol-S0}
    S_0(v) = -\frac{8}{\sqrt{3}} M \sqrt{\Lambda} \,v^2 +\text{const}\,.
\end{equation}
{Inserting} \eqref{eq:7-applic-exp-S} in \eqref{eq:7-eq-exact-S}, the~leading contributions in $\hbar^2$ to $\mathcal{C}_{\Re}$ result in
\begin{equation}
     -\frac{3}{64 M} \left((\partial_v S_0)^2 +2\hbar^2 (\partial_v S_0) \partial_v S_1\right) + 4 M \Lambda v^2 =-\frac{9\hbar^2}{256 M} \frac{(\partial_v^2 S_0)^2}{(\partial_v S_0)^2} +\frac{3\hbar^2}{128 M} \frac{\partial_v^3 S_0}{\partial_v S_0}\,.
\end{equation}
{Making} use of the solution \eqref{eq:7-appl-sol-S0}, for~which $\partial_v^3 S_0 =0$, we arrive at the following expression for $S_1$:
\begin{equation}
    S_1(v) = \frac{1}{256 \sqrt{3\Lambda}} \,v^{-2} + \text{const}\,.
\end{equation}
{Therefore,} the gravitational phase in \eqref{eq:separation-psi-rho-s} at the leading order is:
\begin{gather}
    S(v) = -\frac{4}{\sqrt{3}} M \sqrt{\Lambda}\, v^2 +\frac{\hbar^2}{256 \sqrt{3 \Lambda}}\, v^{-2} + \text{const}\,.\label{eq:7-sol-S}
\end{gather}

We remark that the presence of a non-zero cosmological constant $\Lambda$ is compatible with a de Sitter phase: indeed, for $\Lambda=0$, we would not have a viable cosmological solution of Einstein's equations.
If one considered Equation~\eqref{eq:7-applic-WDW-v} with $\Lambda=0$, the~interpretation would be non-trivial: this case would correspond to a vacuum universe having from \eqref{eq:7-appl-HJ} $S_0=0$ at the classical level (actually $S_0(v)=const$, which can be put to zero) but~with total phase $S(v) = \hbar^2 S_1(v)$. Therefore, the model would yield a ``purely quantum'' trajectory without the classical background. 
For this reason, in~the following, we always consider $\Lambda>0$.

The Bohmian trajectory for $v$ can now be inferred from the guidance equation
\begin{equation}\label{eq:guidance-coord-time}
    \dot{v} =\frac{dv}{dt}= -\frac{3}{32M} \dev{S}\,.
\end{equation}
{Using} the phase computed in \eqref{eq:7-sol-S}, one obtains {the analytical solution}
\begin{equation}\label{eq:7sol-v-guidance-coord-time}
    v(t) = \frac{\sqrt{2}}{(3\Lambda)^{1/8}} \left( e^{\sqrt{3\Lambda} (t-t_0)} -\frac{\sqrt{3}\hbar^2}{16\cdot 256 M \sqrt{\Lambda}} \right)^{1/4}\,,
\end{equation}which is valid in a limited range $t_0 <t_{min}< t < t_{max}$ inside the de Sitter phase (where $t_0$ is the beginning of such phase). 
{For a detailed discussion on computational techniques to solve the guidance equation, we instead refer the reader to~\cite{bib:oriols-2019-bohmian}.}
By definition of $v$, the~classical regime would correspond to 
\begin{equation}\label{eq:7-sol-v-t}
    v_0(t) = e^{\frac{\sqrt{3\Lambda}}{4} (t-t_0)} = a_0(t)^{\frac{3}{2}} \equiv \left(e^{\Hubble (t-t_0)}\right)^{3/2}\,,
\end{equation}where $a_0(t)$ is the classical solution of the de Sitter phase. For~simplicity, we reabsorb the numerical factor in front inside $t_0$; we also consider $t_0=0$, effectively putting the origin of the coordinate time at the beginning of the de Sitter phase such that $\Hubble = \sqrt{3\Lambda}/6$. 

For the following analysis, it is more convenient to work in the conformal time gauge $N=a = v^{\frac{2}{3}}$, in~which the classical scale factor takes the form $a(\eta) = -1/(\Hubble \eta)$. The~dBB trajectory \eqref{eq:7sol-v-guidance-coord-time} now becomes
\begin{equation}\label{eq:7sol-v-guidance-conf-time}
    v(\eta) = \left(\frac{2}{3\Hubble}\right)^{\frac{1}{4}} \left[ \left(-\frac{1}{\Hubble\eta}\right)^6 -\frac{\hbar^2}{32\cdot 256 \,\Hubble}\right]^{\frac{1}{4}}\,.
\end{equation}
{Let} us remark that in our analysis, the de Sitter model does not correspond to an eternal inflation scenario; therefore, the solution above is valid inside an interval $\eta_i < \eta< \eta_f$, analogous to \eqref{eq:7-sol-v-t}.

\begin{figure}[h]
	\centering
	\includegraphics[width=0.7\linewidth]{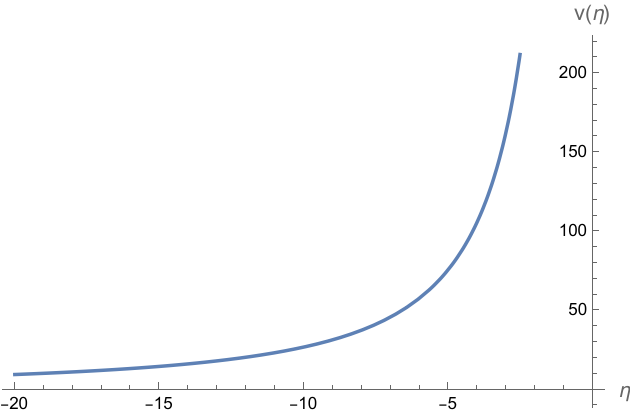}
	\caption[dBB trajectory of the volume variable]{Plot of the computed dBB trajectory \eqref{eq:7sol-v-guidance-conf-time} in conformal time. As~an effect of the quantum correction, the~variable remains above the $\eta$ axis and does not reach zero. We use as reference values $\Hubble=0.02$ and $\hbar=0.001$ {to properly describe the perturbative regime}.
	\label{fig:7-dbb-trajectory}}
\end{figure}

We observe that in both solutions \eqref{eq:7sol-v-guidance-coord-time} and \eqref{eq:7sol-v-guidance-conf-time}, the~action of the quantum potential results in a ``volume'' (actually its square root), which is non-vanishing in the considered range due to the contribution of $\mathcal{O}(\hbar^2)$ (see Figure~\ref{fig:7-dbb-trajectory}), striking a difference from the classical solution. Indeed, Equation~\eqref{eq:7sol-v-guidance-conf-time} vanishes for $\eta^* = - 2\sqrt{2}\Hubble M^{1/6}/\hbar^{1/3}$, which falls outside our de Sitter~approximation. 

\section{A Born--Oppenheimer-like Separation of the~System}\label{sec:dbb-bo}

The evolution of the primordial universe is deeply tied to the matter content. However, one can investigate the dynamics of the full system by implementing a Born--Oppenheimer-like separation of the two components, namely geometry and matter, in~view of the difference in their typical scales. 
The ``lighter'' matter content is described as a fast quantum contribution, while the gravitational sector is heavier (due to the Planck scale being several orders of magnitude higher) and therefore slow. 
In other words, one could think of the total wave function as a separable entity $\Psi(g,m) = \psi(g) \chi(m,g)$, $g$ being the gravitational degrees of freedom and $m$ the matter ones as~in the exact factorization program of molecular physics~\cite{bib:abedi-gross-2010-factor,bib:agostini-gross-2021-factor}.
This line of research has been extensively applied in the context of Canonical Quantum Gravity and Quantum Cosmology and, through perturbative expansions, it has led to the investigation of quantum-gravitational modifications to the primordial matter evolution, even at the level of the inflaton field, therefore influencing the primordial perturbation spectrum~\cite{bib:brizuela-kiefer-2016-desitter,bib:brizuela-kiefer-2016-slow-roll,bib:bortolotti-2022,bib:maniccia-montani-torcellini-2023}.

Therefore, we now introduce the small perturbations of the inflaton sector, represented by a single scalar field. The~gauge-invariant formulation of such fluctuations allows us to write the corresponding super-Hamiltonian as
\begin{equation}\label{eq:H-phi-quantized}
    \hat{H}^{\phi} = \frac{1}{2 v^{2/3}} \sum_{\mathbf{k}} \left(-\hbar^2 \deusq{} + \omega_k^2(\eta) \uscal^2\right)\,,
\end{equation}$\varphi_{\mathbf{k}} = a \delta \phi_{\mathbf{k}}$ being the associated Mukhanov--Sasaki variable~\cite{bib:weinberg-grav-cosm,bib:mukhanov-1985}, i.e.,~the Fourier component of the scalar field perturbation recasted via the universe scale factor. 
We stress that the inflaton field  propagates on a background which is not exactly the classical de Sitter one, due to the Bohmian modifications to the scale factor trajectory computed in the previous~section.

We now employ the Born--Oppenheimer separation between the gravitational and matter components: $\Psi(v,\varphi) = \psi(v) \chi(\varphi,v)$. 
Here, the inflaton field is understood as a purely quantum component, while for the gravitational sector, we wish to use the dBB picture; the ultimate goal is to infer the inflaton evolution with small quantum corrections coming from the gravitational background. Therefore, keeping in mind the form \eqref{eq:7-dbb-ansatz}, we start with the following ansatz:
\begin{equation}\label{eq:7-appl-ansatz-bo}
    \Psi(v,\varphi) = \ampl(v)\, e^{\frac{i}{\hbar}S(v)}\, \chi(\varphi,v)
\end{equation}
{As} previously noted, this ansatz is similar to Vilenkin's one~\cite{bib:vilenkin-1989}. 
An important deviation from that treatment (and also from the works~\cite{bib:kiefer-1991,bib:bertoni-venturi-1996,bib:brizuela-kiefer-2016-desitter,bib:brizuela-kiefer-2016-slow-roll,bib:montani-digioia-maniccia-2021,bib:maniccia-montani-torcellini-2023}) is that here, we do not implement a full WKB expansion of the wave function. 
Instead, the~form \eqref{eq:7-appl-ansatz-bo} can be interpreted as a WKB expanded gravitational sector with a purely quantum matter~component.

In the spirit of the BO separation, we implement the following~assumptions:
\begin{itemize}
    \item We consider the gravitational constraint to be satisfied {a priori}, similarly to~\cite{bib:vilenkin-1989}. Actually, it can be shown that this requirement is equivalent in the canonical picture to performing an extended BO approximation and averaging over the small graviton fluctuations since the associated gauge reconstructs the gravitational WDW constraint at first order~\cite{bib:maniccia-montani-antonini-2023}.

    \item To characterize the regime in which the gravitational sector is close to the classical behavior, we consider it to have a large momentum~\cite{bib:landau3}. Recalling Equation~\eqref{eq:7-guidance-eq}, this can expressed through its phase by the condition:
    \begin{equation}\label{eq:7-condiz-eikon}
        \partial_v S \gg 1\,.
    \end{equation}
    
    This condition is usually referred to as the \emph{eikonal} approximation~\cite{bib:landau2}, which is a simpler case of the WKB one. In~other words (by employing the hypothesis \eqref{eq:7-condiz-eikon} together with the form \eqref{eq:7-appl-ansatz-bo}), we effectively use a gravitational WKB scheme, considering the small quantum deviations via the dBB~picture.
    
    \item To express the small dependence of the fast matter sector on the volume-like variable, we consider the following adiabatic approximation
    \begin{equation}\label{eq:7-condizderivate}
        \partial_v^2 \chi \ll (\partial_v \psi) \partial_v \chi\,,
    \end{equation}    where the smallness is here expressed with respect to the gravitational wave function contribution. On~the other hand, the~terms $\partial_v \chi$ are the objects of our investigation and will be the source of the relevant first-order corrections to the inflaton dynamics.
\end{itemize}

Putting together the ansatz \eqref{eq:7-appl-ansatz-bo}, which obeys the total Wheeler--DeWitt equation, and~the background description of Section~\ref{sec:dbb-background}, the~following system of quantized constraints holds:
\begin{gather}
    \left(\frac{3\hbar^2}{64 M} \partial_v^2  + 4M \Lambda v^2\right) \ampl \,\phase=0\,,\label{eq:7-appl-gravH}\\
    \left(\frac{3\hbar^2}{64 M} \partial_v^2  + 4M \Lambda v^2 +\hat{H}^{\phi}\right) \ampl\, \phase \chi =0\,,\label{eq:7-appl-totalH} 
\end{gather}
where $\hat{H}^{\phi}$ is given by \eqref{eq:H-phi-quantized}. Now expanding the second expression, we have 
\begin{equation}\label{eq:7-appl-totalH-ext}
    \begin{split} \frac{3\hbar^2}{64M}& \left[ \partial_v^2 \ampl +2\frac{i}{\hbar} (\partial_v \ampl) \partial_v S +\frac{i}{\hbar}\ampl \,\partial_v^2S -\frac{\ampl}{\hbar^2}\left(\partial_v S\right)^2 \right] \phase \chi+ \frac{3}{64M} \ampl \,\phase \partial_v^2 \chi+4M \Lambda v^2\, \ampl \,\phase\,\chi \\
    &+\frac{3}{32 M} \left( \partial_v \ampl\,\phase +\frac{i}{\hbar} \ampl (\partial_v S) \phase\right) (\partial_v \chi) +\frac{1}{2 v^{2/3}} \ampl \,\phase\sum_{\mathbf{k}} \left(-\hbar^2 \deusq{} + \omega_k^2(\eta) \uscal^2\right)\chi =0\,.
    \end{split}
\end{equation}
{This} expression is simplified by the background constraint; see Equations~\eqref{eq:7-applic-real-constr} and \eqref{eq:7-applic-im-constr}. By~the assumptions \eqref{eq:7-condiz-eikon} and \eqref{eq:7-condizderivate}, we truncate all the higher-order contributions, finding
\begin{equation}\label{eq:7-din-chi-sistem}
    \frac{3}{32M} \left(\frac{i}{\hbar}\ampl (\partial_v S) \partial_v \chi \right)\phase+\frac{1}{2 v^{2/3}} \ampl \,\phase\sum_{\mathbf{k}} \left(-\hbar^2 \deusq{} + \omega_k^2(\eta) \uscal^2\right)\chi =0\,.
\end{equation}
{Here}, we stress that only the dominant contribution from the gravitational momentum operator acting on $\Psi$ has been taken into account. The~other terms which are present in the second line of Equation~\eqref{eq:7-appl-totalH-ext} are negligible in the present B-O picture. More specifically, we  omit both $\partial_v^2 \chi$ due to the condition \eqref{eq:7-condizderivate} with respect to $(\partial_v S)\partial_v \chi$, and~the term with $\partial_v \ampl$. 
The motivation for the latter is that the background amplitude is determined by the dBB picture to be
\begin{equation}\label{eq:7-legame-ampl-phase}
    \ampl \propto \frac{1}{|\partial_v S|^{1/2}}\,,
\end{equation}(see Equation~\eqref{eq:7-rho-function-s}), and~therefore subdominant from condition \eqref{eq:7-condiz-eikon}.

To recover a physical description of the inflaton evolution, we have to recover a time parameter from the WDW equation.
It is now straightforward to implement the time definition \emph{à la} Vilenkin~\cite{bib:vilenkin-1989}, up~to a Planckian numerical factor~\cite{bib:montani-digioia-maniccia-2021}: the matter evolution is expressed by its dependence on $S$ via
\begin{equation}\label{eq:7-time-def}
    i\hbar \partial_t \chi \equiv -\frac{3}{32M} \frac{i}{\hbar}(\partial_v S) \partial_v\chi\,.
\end{equation}
{Such a time construction, analogous to the one in}~\cite{bib:kiefer-1991}, has been debated to lead to non-unitary effects in the canonical picture \cite{bib:kiefer-2018,bib:montani-digioia-maniccia-2021};
 however, the~case of investigation here is different since we are using the dBB interpretation instead of the canonical quantization. 
Indeed, here, the modified dynamics for the inflaton sector will be induced by the fact that the gravitational component experiences a non-classical evolution due to the quantum potential, as~is clear in the computation of the dBB trajectory (Section~\ref{sec:dbb-background}). 
Therefore, this approach is not in contrast with the findings of~\cite{bib:maniccia-montani-2022,bib:kramer-chataignier-2021,bib:bertoni-venturi-1996} but~it gives an alternative formulation to study small quantum corrections for the gravity--matter~system.

From Equations~\eqref{eq:7-din-chi-sistem} and \eqref{eq:7-time-def}, the~inflaton field thus experiences a Schr\"odinger-like evolution, behaving as a time-dependent harmonic-oscillator
\begin{equation}
    i\hbar \partial_t \chi = \frac{1}{2 v^{2/3}} \sum_{\mathbf{k}} \left(-\hbar^2 \deusq{} + \omega_k^2(\eta) \uscal^2\right)\chi\,,
\end{equation}
with frequency in Fourier space
\begin{equation}\label{eq:7-tdho-frequency}
    \omega(\eta)^2 = k^2 -\frac{z''}{z}\,
\end{equation}
where $k = |\myvec{k}|$ identifies the mode, $z=a\sqrt{\epsilon}$ being $\epsilon = -\dot{\Hubble}/\Hubble^2$, the first slow-roll~parameter.

Let us summarize our findings until now. By~considering a B-O separation of the scalar inflaton field on top of  the Bohmian de Sitter background of Section~\ref{sec:dbb-background}, with~the clock definition \eqref{eq:7-time-def}, we place ourselves in the same scenario as the cosmological investigations of~\cite{bib:brizuela-kiefer-2016-desitter,bib:brizuela-kiefer-2016-slow-roll,bib:maniccia-montani-torcellini-2023}, with~the fundamental difference that the universe volume experiences the modified Bohmian trajectory \eqref{eq:7sol-v-guidance-conf-time}.

\section{Inflaton Power Spectrum~Analysis}\label{sec:applic-spectrum}

The behavior of the universe volume-like variable computed in Section~\ref{sec:dbb-background} is the cause of a modified frequency of the harmonic oscillator in Equation~\eqref{eq:7-tdho-frequency}; thus, this will result in a deviation from the standard primordial power spectrum. 
To compute these features, we now focus on the behavior of the quantum primordial perturbations on top of the gravitational dBB background of Section~\ref{sec:dbb-background}.

\subsection{Quantum Perturbation~Analysis}\label{ssec:perturbations}

Inspecting Equation~\eqref{eq:7-tdho-frequency}, the~variable $z$ is clearly influenced by the dBB trajectory of Section~\ref{sec:dbb-background}, both through $v(\eta)$ and the slow-rolling parameter $\epsilon$, which itself depends on $v(\eta)$ via the Hubble function.
In this pure de Sitter scheme, we implement a constant $\epsilon$ so that from \eqref{eq:7sol-v-guidance-conf-time}, one finds at the leading order in $\hbar$ the following expression:
\begin{equation}\label{eq:7-pulsaz-hbar}
    \omega(\eta)^2 \simeq k^2 -\frac{2}{\eta^2} +\mu \hbar^2 \eta^4\,,
\end{equation}
where $\mu = \frac{7}{32\cdot 256 M} \Hubble^5$ is a numerical factor. Here, again, in~the limit $\hbar \rightarrow 0$, Equation~\eqref{eq:7-pulsaz-hbar} yields the standard result of cosmological perturbation~theory.

To infer the corresponding power spectrum, we recall that the correlation function must be computed on the eigenstates of the oscillator properly satisfying the Bunch--Davies condition~\cite{bib:weinberg-grav-cosm,bib:martin-vennin-peter-2012}. 
Through the Lewis--Riesenfeld method of invariants~\cite{bib:lewis-1967,bib:lewis-1968,bib:lewis-riesenfeld-1969}, this requires to find the solution to the Ermakov equation:
\begin{equation}\label{eq:ermakov}
        \rho_k^{\prime\prime} + \omega_k^2 \rho_k = \frac{1}{\rho_k^3}  \,,      
    \end{equation}
where $^\prime \equiv \partial_{\eta}$, which is then used to compute the invariant
\begin{equation}\label{eq:invariant-I}
        I = \frac{1}{2}\left[\frac{v_{\mathbf{k}}^2}{\rho_k^2} + (\rho_k \pi_{v_{\mathbf{k}}} - \dot{\rho}_k v_{\mathbf{k}})^2 \right]
    \end{equation}
allowing to determine the oscillator eingenstates (see an analogous implementation in~\cite{bib:maniccia-montani-torcellini-2023}).

However, we must take into account that our frequency \eqref{eq:7-tdho-frequency} has a contribution of order $\hbar^2$. In~an approximate scheme, the~solution to \eqref{eq:ermakov} can also be expanded in such a way that
\begin{gather}
    \omega^2(k,\eta) = \omega_0^2(k,\eta) +\hbar^2 \omega_1^2(k,\eta)\,,\\
    \rho(k,\eta) = \rho_0(k,\eta)  +\hbar^2 \rho_1(k,\eta) \,,
\end{gather}
where we  drop the subscript $k$ for readability.
From comparison with \eqref{eq:7-pulsaz-hbar}, we have that
\begin{equation}\label{eq:7-omega0}
    \omega_0^2(k,\eta) = k^2-\frac{2}{\eta^2}\,,\qquad \omega_1^2(k,\eta) = \mu \eta^4\,.
\end{equation}
{As} a consequence, also the Ermakov equation splits into a contribution of order $\hbar^0$, i.e.,
\begin{equation}\label{eq:7-rho0eq}
    (\rho_0)''+\omega_0^2\, \rho_0 = \frac{1}{(\rho_0)^3}\,,
\end{equation}and next-order contributions. For~the latter, we truncate at the order $\hbar^2$. This is obtained by Taylor expanding the term $1/\rho^3$, which results in:
\begin{equation}\label{eq:7-rho1eq}
    (\rho_1)'' + \omega_0^2\, \rho_1 = -\omega_1^2\, \rho_0 + 3 \frac{\rho_1}{(\rho_0)^4}\,.
\end{equation}
Concerning Equation~\eqref{eq:7-rho0eq}, we immediately recognize that its solution corresponds to the standard result in cosmology, namely,
\begin{equation}\label{eq:7-sol-rho0}
    \rho_0 = \sqrt{\frac{1}{k^3\eta^2} +\frac{1}{k}}\,,
\end{equation}
which already satisfies the Bunch--Davies vacuum~requirement.

For $\rho_1$, we need to solve Equation~\eqref{eq:7-rho1eq}, which is of the second order and inhomogeneous. However, we can carry on a qualitative analysis in the limit of small $\eta$, in~which we aim to compute the spectrum; this regime assures that the inflationary perturbations freeze out~\cite{bib:weinberg-grav-cosm,bib:martin-vennin-peter-2012}. The~term $\omega_0 \rho_1$ with $\omega_0^2 \propto 1/\eta^2$ results in being dominant, while $\omega_1^2 \rho_0 \sim \eta^3$ and $1/(\rho_0)^4 \sim \eta^4$.
Therefore, by~neglecting the last term, we deal with the approximate form of Equation~\eqref{eq:7-rho1eq} in the limit of small $\eta$:
\begin{equation}\label{eq:7-rho1approx}
    (\rho_1)'' +\omega_0^2 \rho_1 = -\omega_1^2 \rho_0\,.
\end{equation}
{Its} general solution is
\begin{equation}\label{eq:7-sol-rho1}
\begin{split}
    \rho_1 (k,\eta) = &\frac{c_1}{\sqrt{k}}\left( \frac{sin(k\eta)}{k\eta} -cos(k\eta)\right) +\frac{c_2}{\sqrt{k}}\left( \frac{cos(k\eta)}{k\eta} +sin(k\eta)\right) \\
    &-\left(\frac{cos(k\eta)}{k^{1/2}}-\frac{sin(k\eta)}{k^{3/2}\eta}\right) \mathcal{I}_1-\left(\frac{cos(k\eta)}{k^{3/2}\eta}+\frac{sin(k\eta)}{k^{1/2}}\right) \mathcal{I}_2\,,
    \end{split}
\end{equation}
where $c_1, c_2$ are numerical constants, and we define the following integrals
\begin{gather}
    \mathcal{I}_1 = \int_1^{\eta} dy\left[ -\frac{\mu\,y^3}{k^{1/2}} \left(\frac{1}{k^3 y^2} +\frac{1}{k}\right)^{1/2} \left( \frac{cos(k\,y)}{k} +y\,sin(k\,y)  \right)  \right] \,,\label{eq:7-integral1}\\
    \mathcal{I}_2 = \int_1^{\eta} dy\left[ \frac{\mu\,y^3}{k^{1/2}} \left(\frac{1}{k^3 y^2} +\frac{1}{k}\right)^{1/2} \left( y\, cos(k\,y)-\frac{sin(k\,y)}{k}  \right)  \right]\,.\label{eq:7-integral2}
\end{gather}
{Here,} we observe that the first line of \eqref{eq:7-sol-rho1} would reconstruct, after~the appropriate Bunch--Davies requirement, the~same $\rho_0$ of the previous order; therefore, we can consider the two coefficients $c_1$ and $c_2$ to be equal to zero such that $\rho_1$ purely describes the quantum correction to the previous order~solution. 

However, for~the final state to determine the Bunch--Davies vacuum, the~function $\rho_1$ must go to zero in the relevant limit. Indeed, in the Bunch--Davies regime, one considers the inflaton wavelength to be small with respect to the curvature (subhorizon) so that $k_{phys}\gg 1$. 
In this way, the~eigenstate should correspond to the (Minkowskian) lowest energy state of the oscillator. For~large values of $k$, it can be shown that \eqref{eq:7-sol-rho1} goes to zero, and~we recall that $\rho_0$ satisfies the requirement by construction; as a consequence, the~total function $\rho = \rho_0 +\hbar^2 \rho_1$ is compatible with the Bunch--Davies~condition.

\subsection{Spectrum~Morphology}\label{ssec:spectrum}

We are now left with the task of computing the power spectrum from $\rho$ given by \eqref{eq:7-sol-rho0} and \eqref{eq:7-sol-rho1}. 
Following the reasoning of~\cite{bib:martin-vennin-peter-2012}, we find that the correlation function provides the contribution $\rho^2(k,\eta)$. Subsequently, the~power spectrum of perturbations stemming from the inflaton field is 
\begin{equation}\label{eq:def-spectrum-limit}
    \mathcal{P}_{\zeta}(k) = \left. \frac{k^3}{4\pi^2} \frac{\rho^2(k,\eta)}{2 a^2 \epsilon} \right|_{-k\eta \ll 1} \,,
\end{equation}
where $\zeta$ is the comoving curvature perturbation, freezing when the perturbations are outside the horizon (i.e., the~super-Hubble limit in Equation~\eqref{eq:def-spectrum-limit}); this variable is related to the Mukhanov--Sasaki one by
\begin{equation}\label{def:curvature-perturbation}
        \zeta = \sqrt{\frac{4\pi G}{\epsilon}}\frac{\varphi}{a}.
    \end{equation}

By considering the lowest-order contribution $\rho_0$, we would clearly recover the standard scale-invariant result, i.e.,~
\begin{equation}
         \mathcal{P}_{\zeta}^{(0)}(k) = \frac{G\,\Hubble_\Lambda^2}{\pi \epsilon} \Bigg|_{k=a\Hubble_{\Lambda}}\,,
    \end{equation}
having valuated the slow-roll parameter $\epsilon$ at the horizon~crossing.

At the leading order, $\rho_1$ gives a small deviation which can be evaluated by performing a series expansion of the integrals in \eqref{eq:7-integral1} and \eqref{eq:7-integral2}. 
Then, computing the expression in $\eta = 2\pi/\tilde{k} \ll 1$, we find
\begin{equation}\label{eq:7-power-spectrum-final}
    \mathcal{P}^{(1)}_{\zeta}(k) = \bar{\mu}\, \hbar^2 \left(\frac{k}{\tilde{k}}\right)^{-4} \left(\frac{\Hubble}{\widetilde{\Hubble}}\right)^{12} \mathscr{P}(k,\tilde{k})\,,
\end{equation}
where $\bar{\mu} \propto 1/M^2$ is a numerical coefficient, $\tilde{k}$ and $\widetilde{\Hubble}$ are reference values for the mode and Hubble parameter in the considered limit, and~$\mathscr{P}$ is the following polynomial function in $k$:
\begin{equation}\label{eq:polinomio}
\begin{split}
    &\mathscr{P} (k, \tilde{k}) = \tilde{k}^{-8} \left[ 20 \pi^6 k^6 (90-90 k^2-35 k^4+9 k^6) 
    +1152 \pi^5 \tilde{k} \,k^4 (5+3 k^2) \right.\\
    &\quad-90 \pi^4 \tilde{k}^2\,k^4 (90-90 k^2-35 k^4+9 k^6)  -2880 \pi^3 \tilde{k}^3\,k^2 (5+3 k^2)\\
    &\quad\left. +90\pi^2 \tilde{k}^4\,k^2 (90-90 k^2-35 k^4+9 k^6)  + 45 \tilde{k}^6(90-90 k^2-35 k^4+9 k^6) \right]^2.
    \end{split}
\end{equation}
{For the CMB spectrum, a~standard value of} $\tilde{k}$ is the pivot scale $\tilde{k} \simeq 0.002\,\text{M\,pc}^{-1}$; see, for example, \cite{bib:baumann-2009}. 

\begin{figure}[h]
	\centering
	\includegraphics[width=0.7\linewidth]{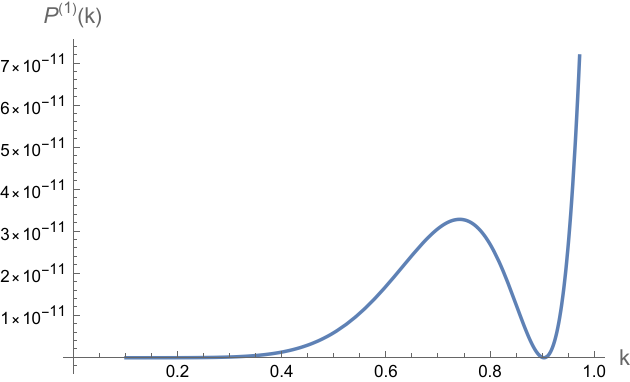}
	\caption[dBB power spectrum of primordial perturbations]{Plot of the computed dBB power spectrum \eqref{eq:7-power-spectrum-final} with reference values $\tilde{\Hubble}=2$, $\Hubble=0.02$, $\tilde{k}=0.027$, {where we have implemented Planckian-like units $\hbar=0.001$, $M=10$, $c=1$, which correspond to setting to one the quantity $10^5/(1.22 \cdot2^{13} M)$, appearing in the coefficient $\mu$ of modified frequency \eqref{eq:7-omega0}. In~this setting, $\hbar$ behaves as the small expansion parameter, and the pivot scale $0.002\,\text{M\,pc}^{-1}$ corresponds to the value $\tilde{k} = 0.027$ suitable to ensure an accurate mathematical integration.}
	\label{fig:7-dbb-power-spectrum}}
\end{figure}

A plot of the obtained power spectrum is provided in Figure~\ref{fig:7-dbb-power-spectrum}.
The obtained correction is clearly not scale invariant; moreover, we observe the presence of a minimum at a point much larger than the pivot scale $\tilde{k}$. 
The presence of the numerical factors $\hbar^2$ and $1/M^2$ in front assures that $\mathcal{P}^{(1)}_{\zeta}$ constitutes a small deviation from the standard~result.

Let us now discuss additional quantities describing the behavior of this primordial signature. The~scale dependence of the primordial power spectrum is identified via its spectral index $n_s$, defined as
\begin{equation}\label{eq:def-ns}
    \mathcal{P}(k) = A\,k^{n_s-1}\,,
    \end{equation}
which therefore can be obtained by 
\begin{equation}
        n_s = 1+ \frac{\mathrm{d}\,ln \,(\mathcal{P}(k)/A)}{\mathrm{d} \,ln\, k}\,.
    \end{equation}
{Likewise}, one can define as additional observables the so-called running $\alpha_s$
\begin{equation}\label{eq:def-as}
     \alpha_s = \frac{\mathrm{d}\,n_s}{\mathrm{d} \,ln\, k}\,
    \end{equation}
and the running of the running $\beta_s$
\begin{equation}\label{eq:def-bs}
    \beta_s = \frac{\mathrm{d}\,\alpha_s}{\mathrm{d} \,ln\, k}\,.
    \end{equation}
{It} is understood that these quantities are evaluated at the horizon exit, i.e.,~at the pivot scale $k = \tilde{k}$ implemented previously. The~dependence of such parameters on the number of e-folds is influenced by the specific model of inflation; see, for example, the study~\cite{bib:garciabellido-2014}. 

\begin{figure}[h]
	\centering	
	\includegraphics[width=0.7\linewidth]{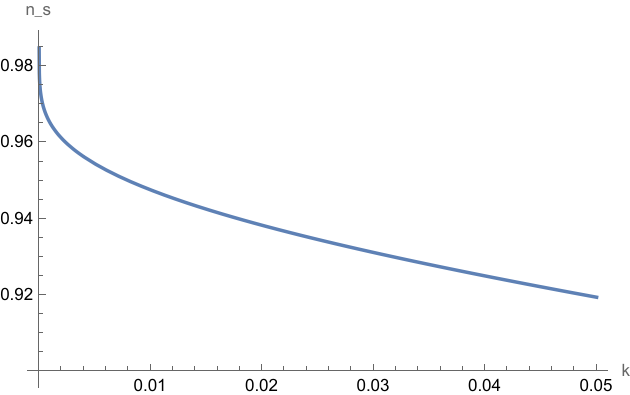}
	\caption[dBB spectral index of primordial spectrum]{Plot of the spectral index $n_s$ corresponding to the power spectrum $\mathcal{P}_{\zeta}^{(1)}$ in \eqref{eq:7-power-spectrum-final}. Reference values are $\tilde{\Hubble}=2$, $\Hubble=0.02$, $\tilde{k}=0.027$ in Planckian-like units $\hbar=0.001$, $M=10$, $c=1$, {see also the caption of Figure~\ref{fig:7-dbb-power-spectrum}}.
	\label{fig:ns}}
\end{figure}

\begin{figure}[h]
	\centering
	\includegraphics[width=0.8\linewidth]{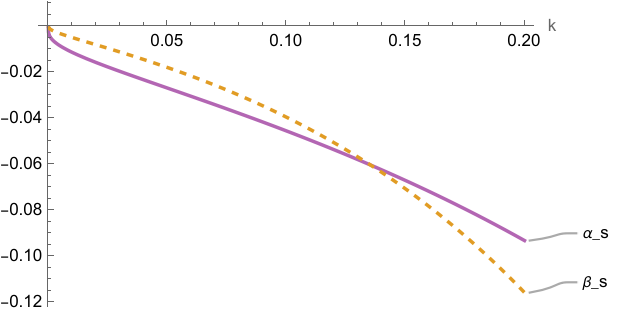}
	\caption{Behavior of the runnings $\alpha_s$ an $\beta_s$ corresponding to the power spectrum $\mathcal{P}_{\zeta}^{(1)}$ in \eqref{eq:7-power-spectrum-final}. Reference values $\tilde{k}=0.002$, $\tilde{\Hubble}=2$, $\Hubble=0.02$, $\hbar=0.001$, $M=10$ in Planckian-like units ($\hbar=0.001$, $M=10$, $c=1$).
		\label{fig:runnings}}
\end{figure}

The interesting case stems from the result $\mathcal{P}_{\zeta}^{(1)}$, where the dependence on the scale $k$ is expressed by the polynomial in \eqref{eq:polinomio}. 
Here, we expect the spectral index to depart from unity; computing the corresponding value from the total power spectrum, we now have a polynomial function of $k$, which is plotted in Figure~\ref{fig:ns}.
As a consequence, we now have non-trivial runnings $\alpha_s$ and $\beta_s$, corresponding again to polynomial functions in the wave number $k$. A~comparison of the two functions provided in Figure~\ref{fig:runnings} shows that, while both parameters are compatible with zero in the limit of small $k$, they actually invert their behavior around $k^* \simeq 0.13$, a~value approximately five times larger than the pivot scale here~considered. 

Such behavior of the two runnings, also referred to as inversion of the hierarchy, is deduced by observations of the PLANCK satellite; see~\cite{bib:planck-results-2018}.
We stress that here, the~deviation of the spectral index and the associated runnings are a consequence of the small dBB corrections alone (their smallness is assured by the presence of the terms $\hbar^2/M^2$ in \eqref{eq:7-power-spectrum-final}) due to the simplifications of the model applied. 
In this sense, they could be overcome by greater-order effects, for~example, by taking into account the more refined slow-roll approximation, or~via other models of inflation. Nonetheless, this prediction of the present perturbative model provides interesting insights for future~investigations.

\section{Discussion and~Outlooks}\label{sec:conclusions}

In the analysis above, we implemented the dBB approach to describe the quantum gravity correction to a standard inflationary spectrum.
We started from the Wheeler--DeWitt equation associated to an isotropic universe in the presence of a cosmological constant and a free massless scalar field, according to an inflation-like scenario (the cosmological constant term must be interpreted as the vacuum energy during the slow-rolling phase~\cite{bib:kolb-turner,bib:peebles-cosmology,bib:imponente-montani-2003}) {in an exact de Sitter phase}.

The decoupling between the classical gravity, its quantum corrections, and the pure quantum Hamiltonian governing the scalar field fluctuations has been performed according to a Born--Oppenheimer separation procedure, in~which the quasi-classical gravity is the ``slow''-component, while the quantum matter is the ``fast'' one.
As a result, we have studied the dynamics of a quantum scalar field fluctuation on a classical-like background, constructed via a dBB recasting of the quantum gravitational constraint as a modified Hamilton--Jacobi equation, solved up to order $\hbar^2$. 
This scenario is associated with a correspondingly modified frequency function for the time-dependent harmonic oscillator, which describes the behavior of
the Fourier component of the scalar fluctuations~\cite{bib:martin-vennin-peter-2012,bib:brizuela-kiefer-2016-desitter}.
Thus, we could provide an expression for the inflationary spectrum, when the slow-rolling parameter is taken as constant, finding the $\hbar^2$ corrections with respect to the scale-invariant spectrum of the standard formulation~\cite{bib:weinberg-grav-cosm}.
These outcoming corrections, which primarily affect the gravitational sector and then the inflaton one, have a non-trivial profile in terms of a polynomial representation, exhibiting an oscillating-like behavior; therefore, they alter the entire spectrum in a non-factorizable manner, preventing the scale-invariant result. 
While these deviations diminish as $k$ decreases, they remain significant for large $k$; see Figure~\ref{fig:7-dbb-power-spectrum}.
This aspect opens intriguing scenarios in view of the Planck collaboration results, discussed in~\cite{bib:carsten-2016}. 
We remark that such a result is in contrast with the findings of~\cite{bib:maniccia-montani-torcellini-2023}, where the quantum corrections (computed without the dBB interpretation) resulted in a time-dependent phase and canceled out when computing the spectrum, { and it is also significantly different from the result in~\cite{bib:brizuela-kiefer-2016-desitter} obtained in a WKB treatment without the graviton averaging procedure.} 

Consequently, our interpretation of a gravity--matter system within the dBB framework demonstrates that relevant quantum effects can occur at the leading order for the power spectrum of primordial perturbations. 
This offers an intriguing opportunity to compare the canonical and dBB paradigms within the confines of the Born--Oppenheimer approximation's validity.
However, it must be noted that the possibility of a direct observation of the spectrum corrections here predicted is not a realistic one in the~future.

The result we obtained here for an exact de Sitter phase of the universe could be also generalized to a more realistic inflationary theory, in~which the details of the scalar field potential can play a physical role~\cite{bib:weinberg-grav-cosm}. 
One could take into account the dependence of the slow-roll parameter $\epsilon$ on the quantum trajectory for the volume-like variable $v$, so extending beyond the scope of the pure de Sitter phase ($\epsilon = \text{const}$), which was implemented in~\cite{bib:brizuela-kiefer-2016-desitter,bib:maniccia-montani-torcellini-2023}. 

Another interesting setup would be to consider the primordial perturbations in a Planckian epoch, i.e.,~a potential-free inflaton field: at Planckian temperature, the~most common inflaton potentials~\cite{bib:kolb-turner}
are negligible with respect to its kinetic energy. 
This would require to analyze the HJ Equation \eqref{eq:7-appl-HJ} in absence of a cosmological constant, then use the ``purely quantum'' background (we recall that in the FLRW model, the classical HJ solution requires a cosmological term) to evaluate the spectrum of the free inflaton field. 
However, it must be noted that its observability could have been forbidden in practice by the subsequent inflationary scenario; for a discussion of the possible implications of the primordial spectrum on the inflationary scenario, see~\cite{bib:ashtekar-2010}.

This proposed scenario could be, in principle, applied to any gravity--matter model.
Indeed, the~gravity--matter separation reduces the impact of the quantum gravity correction to a modified HJ dynamics, containing $\hbar^2$ terms; this does not lead to a single closed equation for the HJ function like Equation~\eqref{eq:7-eq-exact-S} but~to two equations for the phase and the amplitude of the wave functional, which are intrinsically coupled. 
The difficulty in treating the generic gravitational picture stands in its functional character: both the real and imaginary part of the dynamics naturally have a functional structure, leading to revised equations in the full superspace~\cite{bib:mtw-gravitation} of the possible 3d geometries (we emphasize that, formally, the supermomentum constraint is also shifted into a modulus and phase). 
The spatial inhomogeneity of the considered gravitational field would essentially result into a 3d-curvature term in the real component of the WDW equation, i.e.,~a potential-like contribution for the HJ functional.
This puzzling feature can be successfully addressed only under a decomposition into independent minisuperspaces, like for the Bianchi universes~\cite{bib:montani-primordialcosmology} or for the generic cosmological solution in the limit of validity of the so-called 
``long-wavelength conjecture'' \cite{bib:montani-1995,bib:kirillov-montani-1997,bib:kirillov-1993}.
Similarly, one could develop this analysis for black hole physics in the limit in which a minisuperspace representation can be applied (see, for instance, the case of ``Polymer Black Holes'' \cite{bib:modesto-2008,bib:pugliese-montani-2020}); clearly, in~this case, a $3+1$-Hamiltonian dynamics could only emerge with a regularization of the spatial gradients of the $3$d metric, reduced to an effective potential for the infinite but discrete degrees of freedom, for~example, by means of a Regge Calculus~\cite{bib:regge-1961,bib:collins-regge-calculus-book}.

Despite possible technical difficulties typical of all functional approaches to quantum physics, we have here provided a general algorithm by combining a B-O separation between matter and gravity with the dBB treatment of the quantum gravitational degrees of freedom. 

This introduces a new point of view on the origin and interpretation of the deviation from a pure QFT-induced spectrum. 
In fact, the~presence of quantum gravity corrections in the inflationary spectrum would open a non-trivial debate on their measurement and, hence, interpretation (\cite{bib:maniccia-montani-antonini-2023} proposed an average over the quantum gravity degrees of freedom before a QFT phenomenology can arise).
Here, these questions are essentially overcome simply because the quantum ripples of the metric are restated via a modified ``trajectory'', i.e.,~a modified Einsteinian gravity, on which QFT can live with its natural predictivity:
{de facto, we have reduced the problem to the study of QFT on a classical-like background but~amended for the $\hbar$ corrections due to quantum gravity.

\section*{Acknowledgement}
G. Maniccia thanks the TAsP INFN initiative for~support and the Institut d'Astrophysique de Paris for hospitality.

\printbibliography

\end{document}